\documentclass[fleqn,10pt]{wlscirep}
\usepackage{txfonts}
\usepackage{cases}
\usepackage{enumitem} 
\usepackage{placeins}
\usepackage{siunitx}
\usepackage{textcomp}

\usepackage{algpseudocode}
\usepackage{algorithm}

\usepackage{here}

\usepackage{bm}
\usepackage{graphicx}
\usepackage{float}
\usepackage{amsmath}
\usepackage{url}

\newcommand{\R}{\mathbb{R}}

\newcommand{\Cov}{\operatorname{Cov}}
\newcommand{\Proj}{\operatorname{Proj}}
\newcommand{\argmin}{\operatorname*{arg\,min}}

\def\newblock{\hskip .11em plus .33em minus .07em}

\title{QUBO Sampling for Mixed Binary Quadratic Programming without Continuous Variable Binarization}

\author[1,*]{Taisei Takabayashi}
\author[1,2,3,4]{Masayuki Ohzeki}
\affil[1]{Graduate School of Information Sciences, Tohoku University}
\affil[2]{Department of Physics, Institute of Science Tokyo}
\affil[3]{Research and Education Institute for Semiconductors and Informatics, Kumamoto University}
\affil[4]{Sigma-I Co., Ltd.}
\affil[*]{takabayashi.taisei.t7@dc.tohoku.ac.jp}

\begin{abstract}
Quantum annealing and related combinatorial optimization methods typically accept quadratic unconstrained binary optimization (QUBO) problems as input, whereas many practical models include constraints and continuous variables. Standard QUBO conversions discretize continuous variables, increasing the binary dimension and often making feasible low-energy states harder to sample. 
We develop a Lagrange-multiplier method for a separable class of mixed-binary quadratic programs (MBQPs) without discretizing continuous variables. The method is based on a finite-temperature free-energy formulation and uses QUBO sampling for the binary sector.
At fixed Lagrange multipliers, the continuous sector is integrated out analytically and enters only the multiplier update, leaving a QUBO over the original binary variables. We evaluate the method on the continuous relaxation of the quadratic $p$-median problem. Compared with a penalty-based QUBO formulation, it generates feasible solutions more reliably. At an appropriate inverse temperature, its conditional relative error is comparable to that of local search for small instances and often lower for the larger tested instances. 
In the time-to-target comparisons, the proposed method maintains high target-reached rates for stringent fixed-accuracy targets. For the larger instances within the tested range, it also reaches the prescribed targets faster than a commercial mixed-integer optimization solver.
\end{abstract}

\begin{document}

\flushbottom
\maketitle
\thispagestyle{empty}

\section*{Introduction}
Quantum annealing (QA)~\cite{kadowaki1998quantum} is an optimization approach that encodes an objective function in an Ising Hamiltonian and searches for low-energy states by evolving a quantum system under a time-dependent Hamiltonian, typically from a transverse-field driver. Binary optimization problems are therefore commonly represented as Ising models or, equivalently, as quadratic unconstrained binary optimization (QUBO) problems. QA and other QUBO-based optimization methods~\cite{Goto_2019_SB,Okuyama_MA_2019,Aramon_2019_Fujitsu_DA,farhi2014_qaoa} have been applied to a wide range of domains such as manufacturing~\cite{venturelli2016quantum,Sawamura_JSP2026,Yonaga2022},
logistics~\cite{ohzeki2019control,Haba2022,Quang2025_agv},
transportation~\cite{neukart2017traffic,shikanai_jpsj_2025,
Goto2026_micro_mobility},
wireless communication~\cite{Takabayashi2025},
childcare support~\cite{Matsumoto2026_child_care},
materials discovery~\cite{doi_frontiers_2023},
and machine learning~\cite{goto_jpsj_2025,haba_NBMF2025}. Many optimization models arising in practice, however, contain explicit
constraints and combine binary decisions with continuous quantities~\cite{Kochenberger2014_QUBO_survey,Glover2019_QUBO_formulation, Belotti2013_minp,Bonami2012_minp}.
This mismatch makes the direct use of QUBO solvers difficult for mixed-binary optimization models.

Standard QUBO conversions encode constraints by penalties and discretize
continuous variables, often enlarging and densifying the resulting QUBO.
Decomposition-based hybrid methods provide another way to separate binary
and continuous variables. For example, Benders decomposition has been
combined with QUBO solvers by assigning the binary master problem to the
QUBO solver and treating the continuous subproblem classically~\cite{zhao_benders2022,hong_benders_qubitefficient2025,
yoshihara_benders2026}.
However, accumulated cuts and auxiliary variables must still be represented
in QUBO form. A central challenge is therefore to retain a QUBO over the original binary variables while incorporating constraints and continuous variables through an external update.

The Ohzeki method addresses the constraint-handling part of this challenge
for binary optimization~\cite{Ohzeki2020,Takabayashi2025_ohzeki_method}.
It introduces the constraint functions linearly into an effective
Hamiltonian and updates the associated Lagrange multipliers from sample averages.
Unlike a direct squared-penalty formulation, this procedure introduces each
constraint function only linearly. Thus, if the constraint function is
linear or quadratic in the binary variables, the inner problem remains a
QUBO, whereas squaring a quadratic constraint would generally produce
higher-order terms and require additional reduction steps.

In this work, we extend this framework to constrained mixed-binary problems.
At fixed Lagrange multipliers, the binary and continuous contributions
separate in the finite-temperature partition function. For the separable
mixed-binary quadratic programming (MBQP) class considered here, the continuous contribution is available in closed form, so the inner sampling problem remains a QUBO over the original binary variables without continuous-variable discretization. 
Algorithmically, the method alternates QUBO sampling at fixed multipliers with projected updates of the Lagrange multipliers. The binary contribution to each update is estimated from the QUBO samples, whereas the continuous contribution is evaluated analytically. The QUBO sampling step can be
implemented using QA or another QUBO solver. Thus, the proposed outer multiplier-update method is not itself a quantum-annealing algorithm; it is solver-agnostic and uses QA or another suitable QUBO sampler only for its inner sampling step.

To evaluate the proposed formulation, we adopt the continuous relaxation
of the quadratic $p$-median problem (QpMP)~\cite{adasme_qpmp2024}. QpMP combines binary facility-selection
decisions, assignment variables, and quadratic interactions between selected
facilities. In this relaxation, the assignment variables are continuous,
but the optimal assignment value for each fixed facility set is unchanged.
The model therefore provides a separable MBQP benchmark in which the number
of assignment variables grows with both the numbers of facilities and
demand points.

The experiments use simulated quantum annealing (SQA), a classical method based on a path-integral Monte Carlo representation of a transverse-field
Ising model~\cite{martonak2002quantum}, as the QUBO sampler.

We compare the proposed method with three baselines: a penalty-based QUBO formulation that embeds all assignment variables and constraints into a single QUBO, a local search heuristic over facility subsets, and a time-to-target benchmark using Gurobi~\cite{gurobi}, a commercial general-purpose mixed-integer optimization solver. The results show that removing the assignment variables from the QUBO substantially improves feasible-solution generation. With an appropriate inverse temperature, the proposed method achieves conditional relative errors comparable to those of local search for small instances and often lower than those of local search for larger tested instances. Moreover, for larger instances within the tested range, its runtime is shorter than Gurobi's time to reach the same target objective value. The experiments further reveal a trade-off between solution quality and feasibility as the inverse temperature changes.

\section*{Background}
A QUBO problem is written as
\begin{equation}
E(\bm{x})=\bm{x}^{\top}Q\bm{x},
\quad \bm{x}\in\{0,1\}^{N}.
\end{equation}
The diagonal and off-diagonal entries of $Q$ encode linear biases and
pairwise interactions, respectively, and an affine change of variables maps
this form to an Ising model.
Practical optimization models are not always unconstrained. Consider, for example,
\begin{equation}
\begin{aligned}
\min_{\bm{x}\in\{0,1\}^{N}} \quad
& f_0(\bm{x}),\\
\mathrm{s.t.}\quad
& F_k(\bm{x})=C_k,
\quad k\in\mathcal{E},\\
& F_k(\bm{x})\le C_k,
\quad k\in\mathcal{I},
\end{aligned}
\label{eq:binary_constrained_problem}
\end{equation}
where $\mathcal{E}$ and $\mathcal{I}$ index equality and inequality constraints, respectively. Because QUBO has no explicit constraint syntax, the restrictions in Eq.~\eqref{eq:binary_constrained_problem} must be reflected in the objective before the model can be submitted to a QUBO solver.

The penalty method is the most common approach. For the equality constraints, one may minimize
\begin{equation}
E_\gamma(\bm{x})
=
f_0(\bm{x})
+
\sum_{k\in\mathcal{E}}\frac{\gamma_k}{2}
\left(F_k(\bm{x})-C_k\right)^2,
\label{eq:penalty_equality}
\end{equation}
where $\gamma_k>0$ is a penalty coefficient. A sufficiently large value raises the energy of infeasible configurations. Inequalities can be converted to equalities using nonnegative slack variables and treated in the same way. Since a QUBO solver accepts binary variables only, a continuous slack $s_k\ge0$ must then be approximated, for example, by
\begin{equation}
s_k\simeq \Delta_k\sum_{r=1}^{R_k}2^{r-1}z_{kr},
\quad z_{kr}\in\{0,1\},
\end{equation}
where $\Delta_k$ is the resolution and $R_k$ is the number of bits.
Penalty-based conversion requires coefficient tuning, can distort and
densify the objective landscape, and increases the binary dimension when
slacks or continuous variables are discretized. These limitations motivate
constraint treatments that keep the QUBO subproblem close to the
original discrete search space.

The Ohzeki method offers an alternative constraint treatment~\cite{Ohzeki2020}. In its original form, applying the Hubbard--Stratonovich
transformation~\cite{Hubbard1959,Stratonovich1957} to squared constraint penalties yields, up to normalization
and contour-dependent terms, the effective Hamiltonian
\begin{equation}
H(\bm{x};\bm{\nu})
=
f_0(\bm{x})
+
\sum_{k=1}^{K}\nu_k F_k(\bm{x}).
\label{eq:effective_hamiltonian_binary}
\end{equation}
Here $\nu_k$ acts as a multiplier associated with
$F_k(\bm{x})=C_k$. Because each constraint enters linearly, the inner
problem remains a QUBO when $F_k$ is linear or quadratic.

At inverse temperature $\beta$, the effective Hamiltonian is associated
with the Boltzmann distribution
\begin{equation}
Q_{\bm{\nu}}(\bm{x})
=
\frac{
\exp[-\beta H(\bm{x};\bm{\nu})]
}{
\sum_{\bm{x}'}
\exp[-\beta H(\bm{x}';\bm{\nu})]
}.
\end{equation}
The multipliers are iteratively updated according to
\begin{equation}
\nu_k^{(t+1)}
=
\nu_k^{(t)}
+
\eta_t
\left[
C_k
-
\left\langle F_k(\bm{x})\right\rangle_{Q_{\bm{\nu}^{(t)}}}
\right],
\label{eq:ohzeki_update}
\end{equation}
where $\eta_t>0$ is the step size and $\langle F_k(\bm{x})\rangle_{Q_{\bm{\nu}}}
=\sum_{\bm{x}}Q_{\bm{\nu}}(\bm{x})F_k(\bm{x})$
denotes the expectation value of the constraint function. In practice, this expectation is estimated from samples $\{\bm{x}^{(s)}\}_{s=1}^{S}$ obtained from the QUBO sampler as
\begin{equation}
\left\langle F_k(\bm{x})\right\rangle_{Q_{\bm{\nu}}}
\simeq
\frac{1}{S}
\sum_{s=1}^{S}
F_k(\bm{x}^{(s)}).
\label{eq:sample_estimation}
\end{equation}
Thus, the QUBO solver serves as an inner sampler for the
multiplier-dependent Hamiltonian, while the sampled constraint residuals
drive the outer multiplier update.
Inequalities can be treated through related partition-function
representations~\cite{Takabayashi2025_ohzeki_method}. 

The present work extends this constraint treatment, previously developed
mainly for binary models, to problems that also contain continuous
variables.

\section*{Method}
The present formulation shares the linear multiplier structure of the original Ohzeki method but uses a different derivation. Whereas the original method starts from squared penalties and a Hubbard–Stratonovich transformation, we impose equality constraints through delta functions and a Fourier representation. For the additive mixed-binary models considered here, this representation separates the binary and continuous factors at fixed multipliers; the continuous factor can be evaluated analytically for the MBQP class introduced below.

\subsection*{Finite-temperature formulation}
Consider the following general mixed-binary problem:
\begin{equation}
\begin{aligned}
  \min_{\bm{x},\bm{y}} \quad
  & f_0(\bm{x})+g_0(\bm{y}),\\
  \mathrm{s.t.}\quad
  & F_k(\bm{x})+G_k(\bm{y})=u_k,
  \quad k=1,\ldots,L,\\
  & \bm{x}\in X\subseteq\{0,1\}^{N},
  \quad
  \bm{y}\in Y\subseteq\R^{M}.
\end{aligned}
\label{eq:general_mip}
\end{equation}
For inverse temperature $\beta>0$, introduce the constrained partition function
\begin{equation}
  Z_{\beta}
  =
  \sum_{\bm{x}\in X}
  \int_{Y}d\bm{y}\,
  \exp\{-\beta[f_0(\bm{x})+g_0(\bm{y})]\}
  \prod_{k=1}^{L}
  \delta\left(u_k-F_k(\bm{x})-G_k(\bm{y})\right).
  \label{eq:constrained_partition}
\end{equation}
As $\beta\to\infty$, the weight concentrates on feasible configurations with the smallest objective values. Equation~\eqref{eq:constrained_partition} is thus a statistical-mechanical representation of Eq.~\eqref{eq:general_mip}.

To incorporate the equality constraints, we use the Fourier representation of the delta function
scaled by $\beta$,
$\delta(r)\propto\int d\hat{\lambda}\,\exp(i\beta\hat{\lambda}r)$.
After setting $\lambda=-i\hat{\lambda}$, the auxiliary variables are
introduced along an imaginary contour. Then, we obtain
\begin{equation}
  Z_{\beta}
  \propto
  \int_{(i\R)^{L}}d\bm{\lambda}\,
  \exp\{-\beta\bm{\lambda}^{\top}\bm{u}\}
  Z_x(\bm{\lambda})I_y(\bm{\lambda}),
  \label{eq:partition_lambda_imag}
\end{equation}
where
\begin{equation}
  Z_x(\bm{\lambda})
  =
  \sum_{\bm{x}\in X}
  \exp\{-\beta H_x(\bm{x};\bm{\lambda})\},
  \label{eq:zx_def}
\end{equation}
\begin{equation}
  H_x(\bm{x};\bm{\lambda})
  =
  f_0(\bm{x})
  -
  \bm{\lambda}^{\top}\bm{F}(\bm{x}),
  \label{eq:effective_hamiltonian_general}
\end{equation}
and
\begin{equation}
  I_y(\bm{\lambda})
  =
  \int_Y d\bm{y}\,
  \exp\{-\beta[g_0(\bm{y})-\bm{\lambda}^{\top}\bm{G}(\bm{y})]\}.
  \label{eq:iy_def}
\end{equation}

The contour representation shows that the multipliers enter the binary
and continuous sectors linearly. We define the multiplier-dependent
factor in the integrand as
\begin{equation}
  \widetilde Z_{\beta}(\bm{\lambda})
  =
  \exp\{-\beta\bm{\lambda}^{\top}\bm{u}\}
  Z_x(\bm{\lambda})I_y(\bm{\lambda}).
  \label{eq:lambda_resolved_partition}
\end{equation}
We define the multiplier-dependent free-energy function as
\begin{equation}
  \Phi_{\beta}(\bm{\lambda})
  =
  -\frac{1}{\beta}\log \widetilde Z_{\beta}(\bm{\lambda})
  =
  \bm{\lambda}^{\top}\bm{u}
  -
  \frac{1}{\beta}\log Z_x(\bm{\lambda})
  -
  \frac{1}{\beta}\log I_y(\bm{\lambda}).
  \label{eq:Fbeta_def}
\end{equation}
In the practical update below, we restrict $\bm{\lambda}$ to a real
effective domain $\mathcal{D}\subset\mathbb{R}^{L}$ where
$I_y(\bm{\lambda})$ is finite. For $\bm{\lambda}\in\mathcal{D}$,
the following normalized distributions are well defined.
\begin{equation}
  Q_{\bm{\lambda}}(\bm{x})
  =
  \frac{1}{Z_x(\bm{\lambda})}
  \exp\{-\beta H_x(\bm{x};\bm{\lambda})\},
  \label{eq:Qlambda_def}
\end{equation}
and the continuous distribution
\begin{equation}
  P_{\bm{\lambda}}(\bm{y})
  =
  \frac{1}{I_y(\bm{\lambda})}
  \exp\{-\beta[g_0(\bm{y})-\bm{\lambda}^{\top}\bm{G}(\bm{y})]\}.
  \label{eq:Plambda_def}
\end{equation}

Provided that differentiation under the integral is justified, Eq.~\eqref{eq:Fbeta_def} yields
\begin{equation}
  \frac{\partial\Phi_{\beta}(\bm{\lambda})}{\partial\lambda_k}
  =
  u_k
  -
  \left\langle F_k(\bm{x})\right\rangle_{Q_{\bm{\lambda}}}
  -
  \left\langle G_k(\bm{y})\right\rangle_{P_{\bm{\lambda}}},
  \label{eq:grad_Fbeta_component}
\end{equation}
where $\langle F_k(\bm{x})\rangle_{Q_{\bm{\lambda}}}=\sum_{\bm{x}\in X}Q_{\bm{\lambda}}(\bm{x})F_k(\bm{x})$ and $\langle G_k(\bm{y})\rangle_{P_{\bm{\lambda}}}=\int_{\bm{y}\in Y}P_{\bm{\lambda}}(\bm{y})G_k(\bm{y})\,d\bm{y}$ denote expectation values with respect to the corresponding normalized distributions. We define the expected constraint residual as
\begin{equation}
\bm r(\bm{\lambda})
=
\bm u
-
\left\langle\bm F(\bm{x})\right\rangle_{Q_{\bm{\lambda}}}
-
\left\langle\bm G(\bm{y})\right\rangle_{P_{\bm{\lambda}}}
=
\nabla_{\bm{\lambda}}\Phi_\beta(\bm{\lambda}).
\label{eq:multiplier_residual}
\end{equation}
The multiplier condition is therefore
$\bm r(\bm{\lambda})=\bm 0$.

On the real effective domain, $\Phi_\beta$ is concave in $\bm{\lambda}$, so this stationarity condition is sought as a maximization problem.
We therefore use the projected gradient-ascent update
\begin{equation}
\bm{\lambda}^{(t+1)}
=
\Proj_{\mathcal{D}_{\varepsilon}}
\left[
\bm{\lambda}^{(t)}
+
\alpha_t
\bm r(\bm{\lambda}^{(t)})
\right],
\label{eq:projected_ascent}
\end{equation}
where $\alpha_t>0$ is the step size and $\mathcal{D}_{\varepsilon}\subset\mathcal{D}$ is a closed safety region on which the continuous integral remains well defined and numerically stable.

In addition to finiteness of $I_y(\bm{\lambda})$, Eq.~\eqref{eq:grad_Fbeta_component} requires differentiability of $\log I_y$ and conditions that permit differentiation and integration to be interchanged. Sufficient conditions for differentiating $I_y(\bm{\lambda})$ under the
integral sign are given in Appendix~\ref{app:regularity}. In the MBQP case below, the closed-form expression for $I_y$ makes the gradient explicit.

\subsection*{Mixed-binary quadratic programming case}

Consider the separable MBQP
\begin{equation}
\label{eq:mbqp}
\begin{aligned}
  \min_{\bm{x},\bm{y}} \quad
  & \bm{x}^{\top}Q\bm{x}+\bm{p}^{\top}\bm{y},\\
  \mathrm{s.t.}\quad
  & \bm{x}^{\top}W_k\bm{x}+\bm{V}_k^{\top}\bm{y}=u_k,
  \quad k=1,\ldots,L,\\
  & \bm{x}\in\{0,1\}^{N},
  \quad
  \bm{y}\in\R_{+}^{M}.
\end{aligned}
\end{equation}
This is Eq.~\eqref{eq:general_mip} with $f_0(\bm{x})=\bm{x}^{\top}Q\bm{x},
\ 
g_0(\bm{y})=\bm{p}^{\top}\bm{y},
\ 
F_k(\bm{x})=\bm{x}^{\top}W_k\bm{x},
\ 
G_k(\bm{y})=\bm{V}_k^{\top}\bm{y}$.
For fixed $\bm{\lambda}$, the binary effective Hamiltonian becomes
\begin{equation}
  H_x(\bm{x};\bm{\lambda})
  =
  \bm{x}^{\top}Q\bm{x}
  -
  \sum_{k=1}^{L}\lambda_k\bm{x}^{\top}W_k\bm{x}.
  \label{eq:mbqp_hamiltonian}
\end{equation}
Thus, the binary subproblem remains a QUBO at every iteration. Updating the multipliers is equivalent to changing the matrix $Q(\bm{\lambda}) = Q - \sum_{k=1}^{L}\lambda_k W_k$.

The continuous integral factorizes as
\begin{equation}
  I_y(\bm{\lambda})
  =
  \prod_{j=1}^{M}
  \int_{0}^{\infty}dy_j\,
  \exp\{-\beta d_j(\bm{\lambda})y_j\},
  \label{eq:mbqp_iy_factorized}
\end{equation}
where
\begin{equation}
  d_j(\bm{\lambda})
  =
  p_j
  -
  \sum_{k=1}^{L}\lambda_k V_{kj},
  \label{eq:dj_def}
\end{equation}
and $V_{kj}$ is component $j$ of $\bm{V}_k$.

For each $j$, the integral in Eq.~\eqref{eq:mbqp_iy_factorized} is finite if and only if $d_j(\bm{\lambda}) > 0$. The effective domain is therefore
\begin{equation}
  \mathcal{D}
  =
  \left\{
  \bm{\lambda}\in\R^{L}
  \mid
  d_j(\bm{\lambda})>0,
  \quad j=1,\ldots,M
  \right\}.
  \label{eq:mbqp_domain_lambda}
\end{equation}
Within this region,
\begin{equation}
  I_y(\bm{\lambda})
  =
  \prod_{j=1}^{M}
  \frac{1}{\beta d_j(\bm{\lambda})}.
  \label{eq:mbqp_iy_closed}
\end{equation}
The continuous contribution to constraint $k$ is consequently
\begin{equation}
  \left\langle G_k(\bm{y})\right\rangle_{P_{\bm{\lambda}}}
  =
  \frac{1}{\beta}
  \sum_{j=1}^{M}
  \frac{V_{kj}}{d_j(\bm{\lambda})}.
  \label{eq:mbqp_grad_iy}
\end{equation}
Thus, the $k$-th component of the multiplier residual is
\begin{equation}
r_k(\bm{\lambda})
=
u_k
-\left\langle\bm{x}^{\top}W_k\bm{x}\right\rangle_{Q_{\bm{\lambda}}}
-
\frac{1}{\beta}
\sum_{j=1}^{M}
\frac{V_{kj}}{d_j(\bm{\lambda})}.
\label{eq:mbqp_residual}
\end{equation}
Thus, the QUBO sampler is used only to estimate expectations over the original binary variables. The binary expectation in Eq.~\eqref{eq:mbqp_residual} is estimated from the QUBO samples using the empirical average as in Eq.~\eqref{eq:sample_estimation}, whereas the continuous contribution is evaluated directly from Eq.~\eqref{eq:mbqp_grad_iy}.

For numerical stability, the multipliers are projected onto
\begin{equation}
\mathcal{D}_{\varepsilon}
=
\left\{
\bm{\lambda}\in\R^L
\mid
d_j(\bm{\lambda})\ge\varepsilon,
\quad \forall j
\right\},
\end{equation}
where $\varepsilon > 0$ is a numerical margin. Since
$\mathcal{D}_{\varepsilon}$ is a closed convex polyhedron, the Euclidean
projection can be computed by a convex quadratic program.

\subsection*{Algorithm}
Algorithm~\ref{alg:proposed_mbqp} summarizes the iterative procedure.
Because practical QUBO samplers need not generate independent samples from the exact Boltzmann distribution, the empirical averages are treated as
stochastic approximations to the theoretical expectations.

The local curvature of the multiplier potential can be expressed in terms of covariance
matrices of the binary and continuous constraint quantities, as shown in Appendix~\ref{app:bb_stepsize}.
However, estimating and inverting the full covariance curvature matrix can be unstable
and costly in practice. We therefore use the Barzilai--Borwein (BB) rule~\cite{BARZILAI_BORWEIN1988} as a scalar
secant approximation:
\begin{equation}
\alpha_t
=
-\frac{
(\bm{\lambda}^{(t)}-\bm{\lambda}^{(t-1)})^{\top}
(\bm{r}^{(t)}-\bm{r}^{(t-1)})
}{
\|\bm{r}^{(t)}-\bm{r}^{(t-1)}\|_2^2
}.
\label{eq:BB_stepsize}
\end{equation}
The derivation of Eq.~\eqref{eq:BB_stepsize} and the corresponding projected update are provided in Appendix~\ref{app:bb_stepsize}.

\begin{algorithm}[t]
\caption{Projected gradient ascent for MBQP using a QUBO sampler}
\label{alg:proposed_mbqp}
\small
\begin{algorithmic}[1]
\Require
Problem data $Q$, $\{W_k\}_{k=1}^{L}$, $\bm{p}$,
$\{\bm{V}_k\}_{k=1}^{L}$, $\bm{u}$;
inverse temperature $\beta$;
number of iterations $T$;
number of samples $S$;
margin $\varepsilon$.
\Ensure
A candidate solution $(\bm{x}_{\mathrm{best}},\bm{y}_{\mathrm{best}})$.
\State Initialize $\bm{\lambda}^{(0)}\in\mathcal{D}_{\varepsilon}$.
\State Initialize an empty candidate pool $\mathcal{P}$.
\For{$t=0,1,\ldots,T-1$}
  \State Construct $Q(\bm{\lambda}^{(t)})=Q-\sum_k \lambda_k^{(t)} W_k$.
  \State Obtain $S$ samples using the QUBO sampler.
  \State Add selected samples to the candidate pool $\mathcal P$.
  \State Estimate $\langle \bm{x}^{\top}W_k\bm{x}\rangle$ for all $k$.
  \State Compute $d_j(\bm{\lambda}^{(t)})$ for all $j$.
  \State Form the multiplier residual $\bm{r}^{(t)} = \bm{u} - \langle\bm{F}(\bm{x})\rangle - \langle\bm{G}(\bm{y})\rangle$.
  \State Determine the step size $\alpha_t$.
\State Update $\bm{\lambda}^{(t+1)} = \Proj_{\mathcal{D}_{\varepsilon}} [\bm{\lambda}^{(t)} + \alpha_t\bm{r}^{(t)}]$.
\EndFor
\State Recover continuous variables $\bm{y}$ for candidates in $\mathcal{P}$.
\State Evaluate the original objective and select the best feasible pair.
\State \Return $(\bm{x}_{\mathrm{best}},\bm{y}_{\mathrm{best}})$.
\end{algorithmic}
\end{algorithm}

After the multiplier iterations, the continuous variables are recovered
for each retained binary candidate. For fixed $\bm{x}$, this recovery can
typically be performed by the linear program $
\min_{\bm{y}\ge0}
\left\{
\bm{p}^{\top}\bm{y}
\ \middle|\ 
\bm{V}_k^{\top}\bm{y}
=
u_k-\bm{x}^{\top}W_k\bm{x},
\ \forall k
\right\}$ or, when additional structure is available, by a simpler closed-form procedure.

The inverse temperature $\beta$ plays two roles. First, it controls the concentration of the binary distribution on low-energy states. Second, the identity $d_j(\bm{\lambda}) \langle y_j\rangle_{P_{\bm{\lambda}}} = 1 / \beta$ shows that $1/\beta$ acts as a barrier parameter in the continuous sector. This identity has the form of the perturbed complementarity condition $s_j y_j = \mu$ in a primal-dual interior-point method~\cite{wright1997primal}, with $\mu=1/\beta$.

Increasing $\beta$ therefore sharpens the binary
distribution while allowing the multipliers to approach the boundary
$d_j(\bm{\lambda}) = 0$, motivating the quality--feasibility comparison in the next section.

\section*{Experiments}
The experiments are designed to answer three questions: whether removing
assignment variables from the QUBO improves feasible-solution generation,
whether the resulting feasible solutions can match or improve upon the
objective quality of local search, and how the proposed method compares with
classical alternatives in time to reach a prescribed solution quality,
both within the range where certified optima are available and in a
larger-size regime where optimality certification becomes difficult.

\subsection*{Problem setting}
We evaluate the proposed method using a continuous relaxation of the quadratic $p$-median problem (QpMP)~\cite{adasme_qpmp2024}. Let $\mathcal{I}=\{1,\ldots,n\}$ and $\mathcal{J}=\{1,\ldots,m\}$ denote the sets of candidate facilities and demand points. Binary variable $x_i$ indicates whether facility $i\in\mathcal{I}$ is opened, while $z_{ij}\geq0$ denotes the fraction of the demand at point $j\in\mathcal{J}$ assigned to facility $i$. The model includes quadratic interaction costs between opened facilities and linear assignment costs:
\begin{equation}
\label{eq:qpmp}
\begin{aligned}
  \min_{\bm{x},\bm{z}}\quad
  &
  \sum_{1\le i<\ell\le n}D_{i\ell}x_i x_\ell
  +
  \sum_{i=1}^{n}\sum_{j=1}^{m}c_{ij}z_{ij}, \\
  \mathrm{s.t.}\quad
  &
  \sum_{i=1}^{n}x_i=p, \\
  &
  \sum_{i=1}^{n}z_{ij}=1,
  \quad j=1,\ldots,m, \\
  &
  z_{ij}\le x_i,
  \quad i=1,\ldots,n,\ j=1,\ldots,m, \\
  &
  x_i\in\{0,1\},
  \ z_{ij}\ge0,
  \quad i=1,\ldots,n,\ j=1,\ldots,m .
\end{aligned}
\end{equation}
Here, $D_{i\ell}$ is the joint opening cost for facilities $i$ and
$\ell$, $c_{ij}$ is the assignment cost from demand point $j$ to
facility $i$, and $p$ is the number of opened facilities.

In the standard QpMP, the assignment variables are binary. However, for any fixed facility-selection vector $\bm{x}$, the assignment subproblem is a linear optimization problem over the simplex of opened facilities, and therefore admits an optimal integral extreme point. Consequently, relaxing $z_{ij}$ to nonnegative continuous variables does not change the optimal objective value~\cite{adasme_qpmp2024}.

Introducing nonnegative slacks $s_{ij}$ such that
$z_{ij} + s_{ij} = x_i$ and collecting
$\bm{y} = (\operatorname{vec}(\bm{z}),
\operatorname{vec}(\bm{s}))^\top$
gives an objective quadratic in $\bm{x}$ and linear in $\bm{y}$.
Because $x_i=x_i^2$, all constraints take the separable quadratic--linear
form of Eq.~\eqref{eq:mbqp}.
This representation is particularly favorable for the proposed method
because the QUBO sampler operates only on the $n$ facility-selection
variables, whereas the assignment and slack variables are handled through the analytic continuous contribution.

\subsection*{Compared methods and implementation details}
We compare the proposed method with the following three baselines chosen to isolate
different aspects of the formulation.
\subsubsection*{Proposed method}
We apply Algorithm~\ref{alg:proposed_mbqp} using the
\texttt{SQASampler} implemented in OpenJij~\cite{openjij}. In all reported
experiments, the SQA sampler was called with a Trotter number of $2$, and the
other OpenJij parameters were left at their default values. The same inverse
temperature $\beta$ was used in the SQA sampler and in the analytic
continuous contribution in Eq.~\eqref{eq:mbqp_grad_iy}. For SQA, this parameter should be regarded as a common control parameter used in both the sampler and the analytic continuous contribution, rather than as a verified physical inverse temperature of the sampled distribution.

The projected BB multiplier update used the domain margin $\varepsilon=10^{-6}$,
the initial step size $\alpha_{\mathrm{init}}=10^{-2}$, and the clipping range $\alpha_{\min}=10^{-8},\quad\alpha_{\max}=10^{2}$. The maximum number of multiplier iterations was set to $30$. For QpMP, any sample satisfying $\sum_i x_i=p$ defines a feasible facility subset. During the multiplier iterations, all
distinct samples satisfying this cardinality condition were stored in the
candidate pool. After the iterations, each retained facility subset was
evaluated by the original QpMP objective: each demand point was assigned to
the lowest-cost opened facility, which is optimal for the continuous
assignment subproblem. The best feasible solution in this pooled set was
returned.

\subsubsection*{Penalty QUBO}
The penalty-based QUBO baseline, referred to as penalty QUBO in the figures, treats $z_{ij}$ as binary and includes the facility-count, assignment, and linking constraints through penalties. This baseline corresponds to the standard binary-assignment representation of QpMP encoded directly as the following penalized QUBO:
\begin{equation}
\begin{aligned}
  H_{\mathrm{pen}}(\bm{x},\bm{z}) =
  \sum_{i<\ell}D_{i\ell}x_i x_\ell + \sum_{i,j}c_{ij}z_{ij}
  + A_{\rm pen}\left(\sum_i x_i-p\right)^2 
  + B_{\rm pen}\sum_j\left(\sum_i z_{ij}-1\right)^2 + C_{\rm pen}\sum_{i,j}z_{ij}(1-x_i),
  \label{eq:direct_qubo}
\end{aligned}
\end{equation}
where $A_{\rm pen}, B_{\rm pen}, C_{\rm pen} > 0$ are penalty coefficients. In the experiments, we used
$(A_{\mathrm{pen}},B_{\mathrm{pen}},C_{\mathrm{pen}})=(50,1,26)$.
These values were selected through a preliminary hyperparameter search using Optuna~\cite{akiba2019_optuna}
on randomly generated QpMP instances and were then fixed for all reported penalty-QUBO experiments.
The penalty formulation contains $n+nm$ binary variables and introduces
additional couplings through the squared assignment constraints, whereas the
proposed QUBO contains only the $n$ facility-selection variables. The same
OpenJij SQA sampler was used for this baseline. The penalty-QUBO baseline was sampled with inverse temperature $\beta=1$, Trotter number $2$, and $30000$ reads.

\subsubsection*{Local search}
The local-search baseline uses the random-exchange procedure proposed by Adasme et al.~\cite{adasme_qpmp2024}. Starting from a facility subset of cardinality $p$, it explores neighboring subsets by exchanging opened and closed facilities and evaluates each candidate using its optimal assignment. Because the cardinality constraint is maintained throughout and the assignment is optimized for each subset, the method is feasible by construction. The comparison with the proposed method therefore focuses on objective quality.

\subsubsection*{Gurobi}
Gurobi~\cite{gurobi} was used for two purposes.
First, at problem sizes where optimality could be certified within the prescribed time limit, it was used to obtain reference optima for the relative-error evaluation.
Second, it was used as a general-purpose mixed-integer optimization baseline in the time-to-target experiments described below.
For the latter experiments, Gurobi was configured to prioritize primal-solution discovery rather than optimality certification.

\subsection*{Experimental settings and evaluation metrics}

The number of candidate facilities is $n$, and the number of demand points is $m=\lfloor \rho n\rfloor$, where $\rho=m/n$.
The number of opened facilities is $p=\lfloor0.2n\rfloor$.
For each instance, the assignment costs were generated independently from a uniform distribution on $[0,1]$.
In the implementation, a matrix $\widetilde C\in\mathbb{R}^{m\times n}$ was sampled as $\widetilde C_{ji}\sim U(0,1)$, and the cost in Eq.~\eqref{eq:qpmp} was set to $c_{ij}=\widetilde C_{ji}$.
The facility-interaction coefficients were generated by sampling $D_{i\ell}\sim U(0,100)$ for $i<\ell$.

For the quality and feasibility experiments, $\rho\in\{0.1,0.3,0.5\}$.
We used $n\in\{10,20,30\}$ for the comparison with penalty QUBO and $n\in\{10,20,30,40,50,60\}$ for the comparison with local search.
For each $(n,\rho)$, 100 random instances were generated.
The proposed method obtained $1000$ QUBO samples per multiplier iteration and performed at most $30$ multiplier iterations, giving at most $30000$ samples per instance.
Penalty QUBO used $30000$ SQA samples, and local search was run for $30000$ independent trials per instance.

For the proposed method, we tested $\beta\in\{0.01,0.1,1.0\}$. These three values form a coarse logarithmic sweep over two orders of magnitude, chosen to examine the quality–feasibility trade-off arising from the dual role of $\beta$. Smaller $\beta$ broadens the binary sampling distribution and increases the continuous-sector barrier parameter $1/\beta$, whereas larger $\beta$ concentrates the binary distribution while allowing the multipliers to approach the boundary of the effective domain. The tested values are not intended to represent universally optimal choices; intermediate or instance-dependent values may provide better performance, motivating adaptive selection of $\beta$ in future work.

For these quality and feasibility experiments, all instances were small enough for Gurobi to certify optimality within the 600~s time limit.
Let $f_{\mathrm{grb}}$ denote the certified optimum.
For a feasible output $f$, the relative error is
${|f-f_{\mathrm{grb}}|}/{|f_{\mathrm{grb}}|}$.
Because the proposed method and penalty QUBO may fail to return a feasible
solution, this metric is computed conditionally on success and reported
together with the feasibility rate.

We performed two complementary time-to-target (TTT) experiments.
In the first, the proposed method, local search, and Gurobi were compared
using the same prescribed relative-error targets in the range for which the optimum could be certified.
We used
$n\in\{10,20,30,40,50,60\}$,
$\rho=0.1$,
$p=\lfloor0.2n\rfloor$,
and $\beta=0.1$,
with 100 random instances for each $n$.
For each instance, the target was defined relative to the certified optimum $f^\ast$ by $|f-f^\ast|/|f^\ast|\leq\epsilon$, with $\epsilon\in\{10^{-4},10^{-3},10^{-2},10^{-1}\}$.
The proposed method was terminated when the target was first attained,
subject to a maximum of 30 multiplier iterations.
Local search was terminated at the first target-reaching solution, subject to a maximum of 30000 trials.
Gurobi was terminated through an incumbent callback when the target was
first reached, subject to a maximum runtime of 600~s.
Because not every method reaches every target within its prescribed budget, we report both the conditional TTT over target-reaching runs and the target-reached rate.

The second TTT experiment extends the analysis to a larger problem-size
range, $n\in\{10,30,50,70,90,110\}$.
At the upper end of this range, certified optima are no longer reliably
available from Gurobi within the 600~s time limit.
We again fixed $\rho=0.1$, $p=\lfloor0.2n\rfloor$, and $\beta=0.1$,
and used 20 random instances for each $n$.
Since a common relative-error target cannot be defined without a certified optimum, the feasible objective value returned by the proposed method was used as an instance-specific target,
$f_{\mathrm{target}}$.
The Gurobi TTT was defined as the first time at which its incumbent reached
$f_{\mathrm{target}}$.
Runs that did not reach the target within 600~s were treated as
right-censored.
This experiment therefore compares the time required for Gurobi to match
the objective value attained by the proposed method, rather than comparing
the two methods against a common optimality-based target.

For both TTT experiments, Gurobi was configured with
\texttt{MIPFocus=1} to prioritize primal-solution discovery and
\texttt{Threads=1}.
No warm start from the reference or proposed solution was provided.
The proposed-method runtime includes QUBO sampling, multiplier updates,
candidate-pool evaluation, and assignment recovery.

All numerical experiments were performed on a personal computer equipped with an Apple M3 processor and 16~GB of memory, running macOS 15.3.2 and Python 3.12.0. SQA was performed using OpenJij version 0.9.2, and Gurobi version 12.0.1 was used for the mixed-integer optimization benchmarks. No explicit thread-count restriction was imposed on OpenJij or the Python numerical libraries.

\subsection*{Results}
\begin{figure}[t]
\centering
\includegraphics[width=\textwidth]{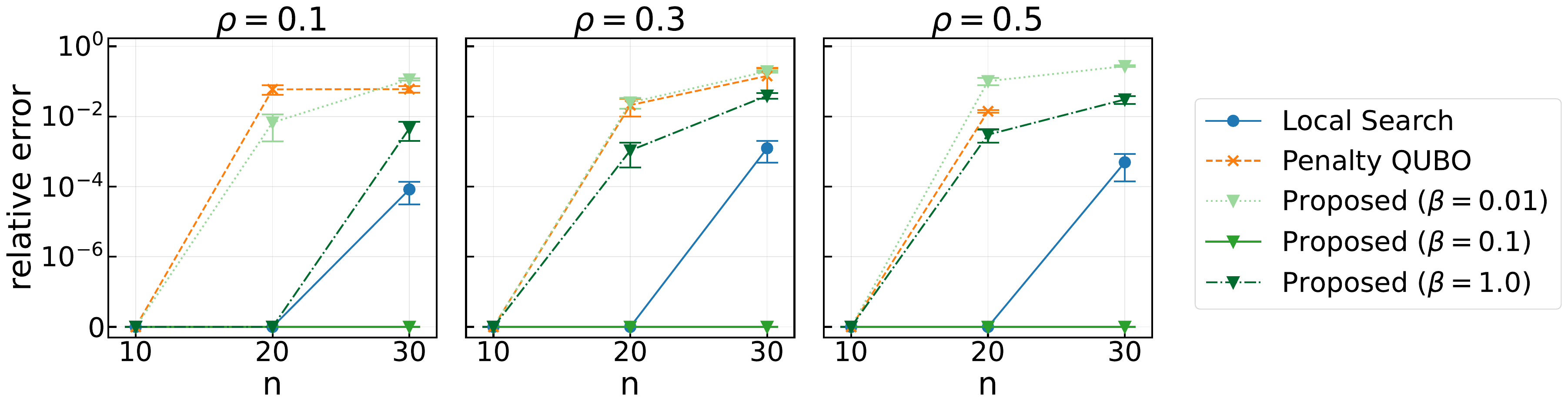}
\caption{Relative error of local search, penalty QUBO, and the proposed method for QpMP instances with $\rho=0.1, 0.3, 0.5$. The proposed method was tested at $\beta=0.01, 0.1, 1.0$. Penalty QUBO embeds the assignment variables and constraints directly in the QUBO. Markers and error bars show the mean and standard error over random instances. 
Relative errors at or below $10^{-8}$ were treated as numerical zero.
For visualization on the logarithmic scale, these zero values are plotted at $10^{-8}$, with the corresponding tick labeled as 0. Downward error bars are omitted when their lower bounds fall at or below the plotting floor. Relative error was computed only for feasible outputs and should be interpreted together with Fig.~\ref{fig:feas_all}. } 
\label{fig:relerr_all}
\end{figure}

\begin{figure}[t]
\centering
\includegraphics[width=\textwidth]{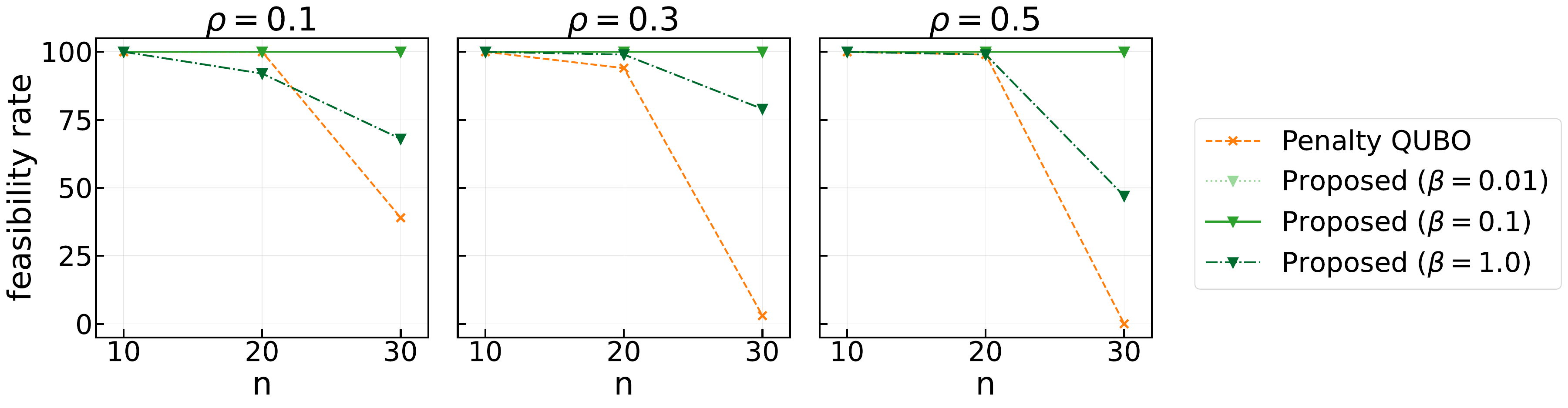}
\caption{Feasibility rate of penalty QUBO and the proposed method for QpMP instances with $\rho=0.1,0.3,0.5$. The proposed method was tested at $\beta=0.01,0.1,1.0$. The feasibility rate is the fraction of instances for which at least one solution satisfying all original QpMP constraints was obtained. Local search is omitted because it is feasible by construction.}
\label{fig:feas_all}
\end{figure}

We first examine the effect of excluding the assignment variables and their constraints from the QUBO. Figures~\ref{fig:relerr_all} and
\ref{fig:feas_all} compare the proposed method with penalty QUBO in terms of conditional relative error and feasibility rate, respectively.

Penalty QUBO produces feasible outputs for small instances at low $\rho$, but its feasibility rate falls rapidly as $n$ or $\rho$ increases. In particular, at $\rho=0.3$ and $\rho=0.5$, some settings near $n=30$ yield almost no feasible solutions.
The decline is consistent with the enlarged search space of the direct
formulation: it contains $n+nm=n+\rho n^2$ binary variables whose
facility-count, assignment, and linking constraints must be satisfied
simultaneously.
By contrast, the proposed method maintains a substantially higher
feasibility rate, particularly at $\beta=0.1$, while achieving
conditional relative errors comparable to or smaller than those of direct
QUBO under the same sampling budget.
Interpreted together, Figs.~\ref{fig:relerr_all} and
\ref{fig:feas_all} show that excluding the assignment variables and their
constraints from the QUBO substantially stabilizes feasible-solution
generation.

\begin{figure}[t]
\centering
\includegraphics[width=\textwidth]{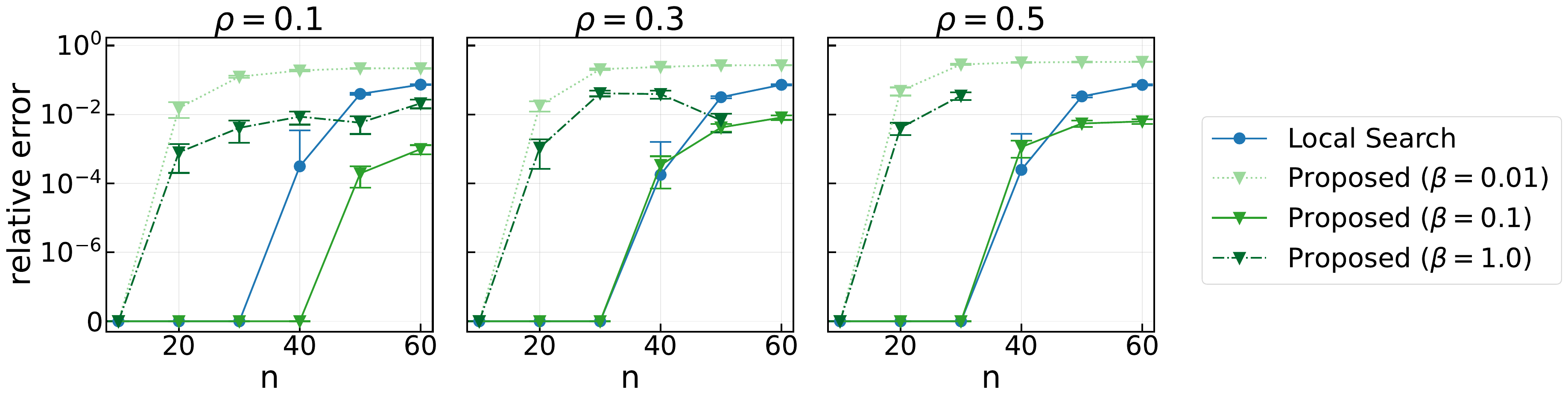}
\caption{Relative error of the proposed method and local search for QpMP instances with $\rho=0.1,0.3,0.5$. The proposed method was tested at $\beta=0.01,0.1,1.0$. For each $(n,\rho)$, the markers and error bars show the mean and standard error over 100 random instances. 
The vertical axis is logarithmic. Relative errors at or below $10^{-8}$ were treated as numerical zero. For visualization on the logarithmic scale, these zero values are plotted at $10^{-8}$, with the corresponding tick labeled as 0. Downward error bars are omitted when their lower bounds fall at or below the plotting floor. Relative error was evaluated only when the corresponding method returned a feasible solution.} 
\label{fig:relerr_ls}
\end{figure}

\begin{figure}[t]
\centering
\includegraphics[width=\textwidth]{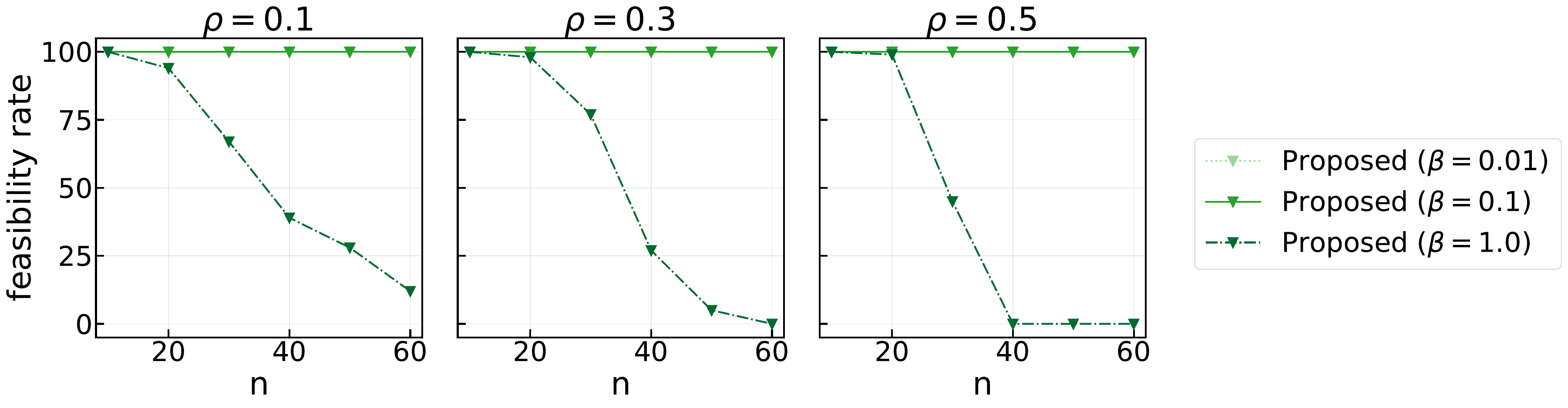}
\caption{Feasibility rate of the proposed method for QpMP instances with $\rho=0.1,0.3,0.5$. The method was tested at $\beta=0.01,0.1,1.0$. Local search is omitted because it directly explores $p$-facility subsets and always constructs a feasible assignment.}
\label{fig:feas_ls}
\end{figure}

Having established the advantage of the reduced QUBO in feasible-solution
generation, we next ask whether the feasible solutions obtained in this way can match or surpass those of a problem-specific classical heuristic in objective quality.
Figures~\ref{fig:relerr_ls} and \ref{fig:feas_ls} show the conditional
relative error and feasibility rate.

The results exhibit a clear quality--feasibility trade-off. At
$\beta=0.01$, feasibility remains high, but the relative error increases
with $n$ and $\rho$. At $\beta=0.1$, the method provides the best
overall balance: it maintains a high feasibility rate and achieves
conditional errors close to those of local search for small instances and
often smaller than those of local search for larger tested instances.
Because local search is feasible by construction, this comparison should be interpreted together with the feasibility rate. Nevertheless, at
$\beta=0.1$, the proposed method maintains high feasibility and yields
lower conditional relative errors than local search in larger tested
instances.

\begin{figure}[t]
\centering
\includegraphics[width=0.85\textwidth]{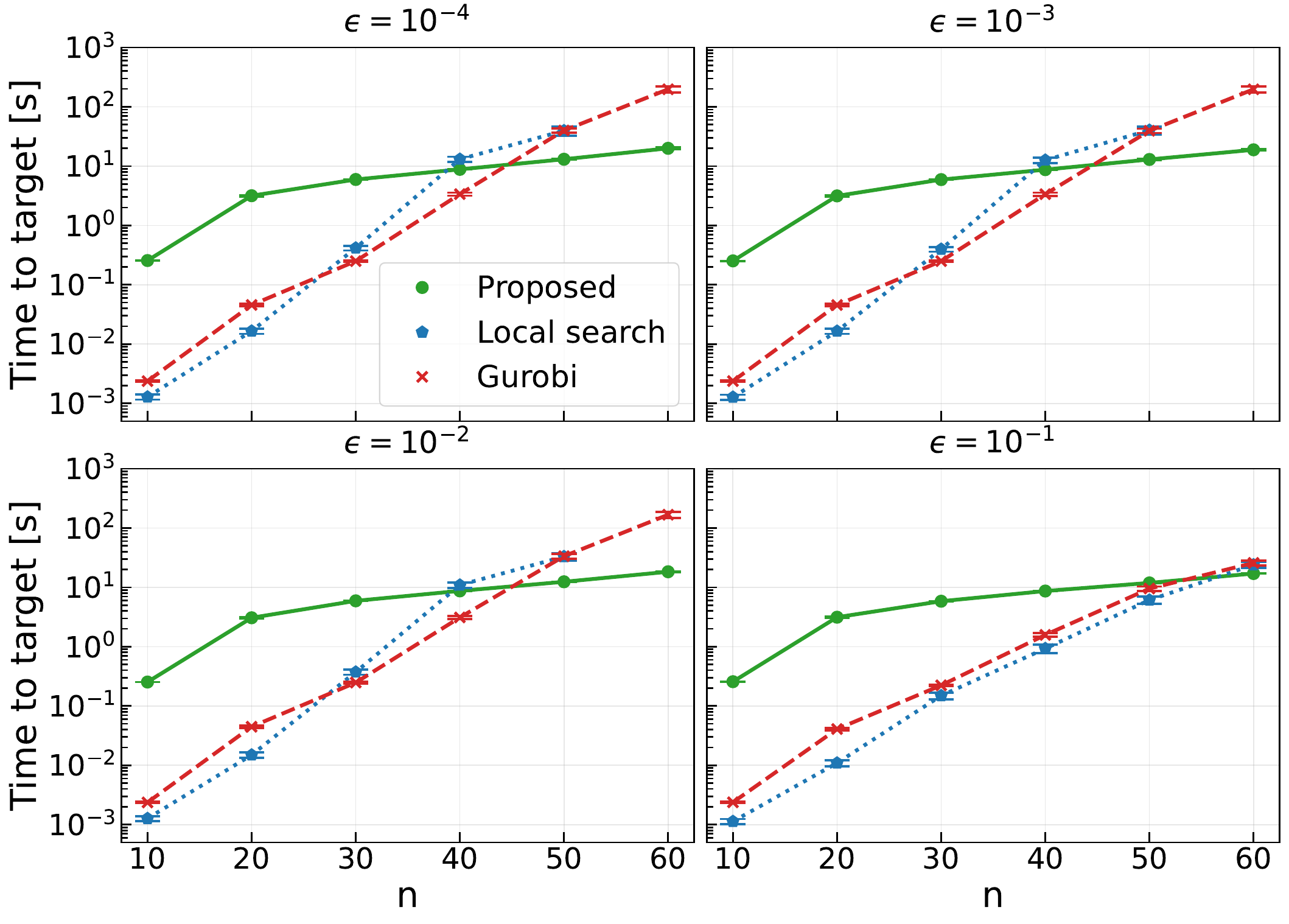}
\caption{
Time to target for the proposed method, local search, and Gurobi for QpMP
instances with $\rho=0.1$ and $p=0.2n$. The target is defined by
$|f-f^\ast|/|f^\ast|\le\epsilon$, where $f^\ast$ is the certified optimum and $\epsilon\in\{10^{-4},10^{-3},10^{-2},10^{-1}\}$.
The proposed method uses $\beta=0.1$.
Markers and error bars show the mean and standard error over runs that reached the target within the prescribed computational budget. Runs that did not reach the target are excluded from this conditional TTT statistic and are reflected in Fig.~\ref{fig:target_reached_rate}.
}
\label{fig:ttt_fixed_error}
\end{figure}

\begin{figure}[t]
\centering
\includegraphics[width=0.85\textwidth]{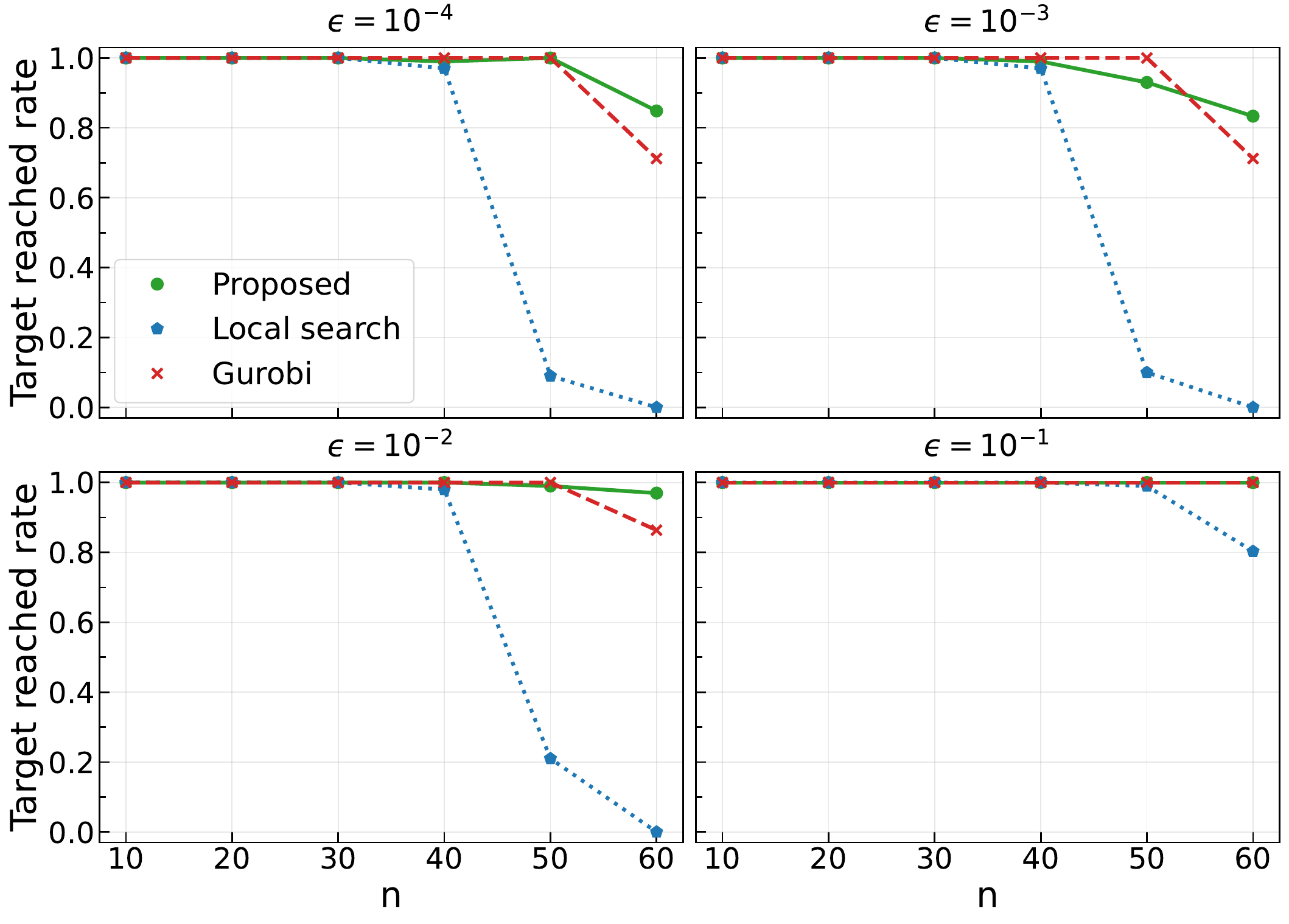}
\caption{
Target-reached rate corresponding to the time-to-target experiment in
Fig.~\ref{fig:ttt_fixed_error}.
For each method, problem size, and target $\epsilon$, the rate is the fraction of the 100 random instances for which the prescribed relative-error target was reached within the corresponding computational budget.
}
\label{fig:target_reached_rate}
\end{figure}

Having compared solution quality and feasibility, we next examine the time required by each method to attain the same prescribed solution accuracy. We first consider the range in which the optimum can be certified by Gurobi, so that the proposed method, local search, and Gurobi can all be evaluated against the same relative-error target. Figures~\ref{fig:ttt_fixed_error} and \ref{fig:target_reached_rate} show the corresponding time to target and target-reached rate, respectively.

As shown in Fig.~\ref{fig:ttt_fixed_error}, local search is the fastest method for the smallest instances, particularly at $n=10$ and $20$. Its conditional TTT, however, increases rapidly with problem size for the more stringent targets $\epsilon=10^{-4},10^{-3}$, and $10^{-2}$. At the same time, Fig.~\ref{fig:target_reached_rate} shows that its target-reached rate remains close to one up to approximately $n=40$ but drops sharply at $n=50$ and reaches zero at $n=60$ for these stringent targets.

In contrast, the proposed method exhibits a more gradual increase in TTT while maintaining a high target-reached rate over the tested range. For the stringent targets, its TTT becomes shorter than that of Gurobi toward the larger problem sizes. For the looser target $\epsilon=10^{-1}$, the differences in both TTT and target-reached rate among the three methods are substantially smaller. Because the reported TTT is conditional on successful target attainment, Figs.~\ref{fig:ttt_fixed_error} and \ref{fig:target_reached_rate} should be interpreted together. A breakdown of the runtime of the proposed method for the representative target $\epsilon=10^{-2}$ is provided in Appendix~\ref{app:runtime_breakdown}.

\begin{figure}[t]
\centering
\includegraphics[width=0.55\textwidth]{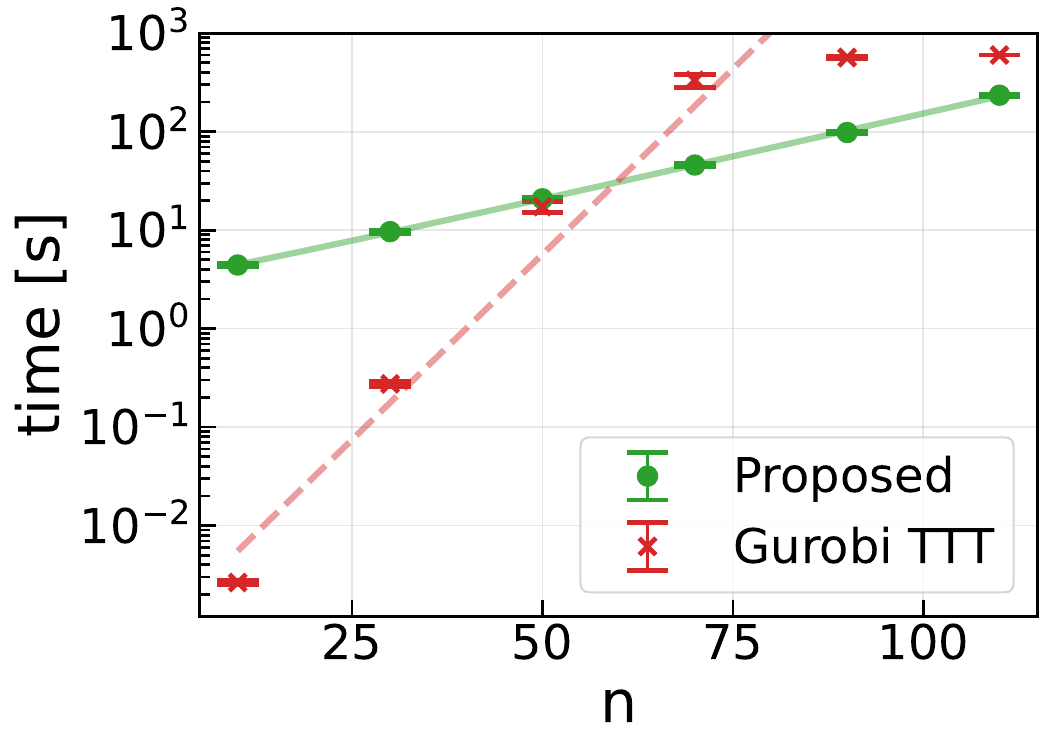}
\caption{
Runtime comparison between the proposed method and Gurobi for QpMP instances with $\rho=0.1$ and $p=0.2n$ over the larger tested problem-size range. The proposed method uses $\beta=0.1$.
For each instance, the target is the feasible objective value obtained by the proposed method, and the Gurobi time to target is the first time at which its incumbent reaches the same value.
Markers and error bars show the mean and standard error over 20 random instances, and the semi-transparent curves show empirical finite-size fits. The Gurobi fit uses only runs that reached the target within 600~s.
}
\label{fig:runtime_scaling}
\end{figure}

The comparisons in Figs.~\ref{fig:ttt_fixed_error} and
\ref{fig:target_reached_rate} require a certified optimum and are therefore restricted to moderate problem sizes. We next extend the analysis to $n\in\{10,30,50,70,90,110\}$. At the upper end of this range, optimality can no longer be certified reliably by Gurobi within the 600~s time limit.
For each instance in this range, we use the feasible objective value obtained by the proposed method as the target and measure the time required for the Gurobi incumbent to reach the same value. Figure~\ref{fig:runtime_scaling} shows this comparison.

As shown in Fig.~\ref{fig:runtime_scaling}, the proposed-method runtime increases gradually over the tested range and is described by $T_{\rm PR}(n)=2.70\exp(0.0403n)+0.413$. The corresponding Gurobi time to reach the same attained objective value increases more rapidly; a fit to the runs that reached the target within 600~s gives $T_{\rm RG}(n)=9.75\times10^{-4}\exp(0.173n)$. Accordingly, Gurobi is faster for the smaller instances, whereas the proposed method becomes faster toward the larger end of the tested range. The fitted curves are empirical finite-size descriptions under the specified computational budgets and are not intended as asymptotic complexity estimates. Nevertheless, the comparison indicates that, over the tested larger-size regime, the computational effort required for Gurobi to match the solution quality attained by the proposed method grows substantially more rapidly. 

\FloatBarrier
\section*{Conclusion and Discussion}
We developed a finite-temperature formulation for a separable class of MBQP in which the continuous sector is integrated out analytically at fixed multipliers. The resulting sampling problem remains a QUBO over the original binary variables, while the continuous contribution is incorporated into the multiplier update in closed form. On the continuous relaxation of QpMP, the method generated feasible solutions more reliably than penalty QUBO. At $\ \beta=0.1$, it maintained high feasibility and achieved conditional relative errors comparable to those of local search at small $n$ and often lower at larger $n$. 
In the time-to-target comparison with fixed relative error, the proposed method maintained a high target-reached rate and became faster than Gurobi for stringent targets toward the larger end of the certified-optimum regime. In the complementary comparison at larger problem sizes, the time required for Gurobi to match the objective value attained by the proposed method increased more rapidly over the tested range.

The results also highlight the dual role of inverse temperature: it controls both concentration on low-energy binary states and the effective barrier in the continuous sector. This coupling explains the observed quality--feasibility trade-off and motivates adaptive temperature control.

Future work should first examine the formulation with physical quantum
annealers and other QUBO samplers, whose outputs may deviate from ideal
Boltzmann samples because of finite-temperature and hardware effects.
Relevant directions include temperature estimation, Gibbs-like sampling for QA~\cite{Benedetti_2016_temeratureQA,Nelson_2022_gibbs_QA}, as well as evaluations using the quantum approximate optimization algorithm
~\cite{Diez_2023_pseudo_boltzmann} and simulated bifurcation
~\cite{kubo_2025_SB_Boltzmann}.

A statistical-mechanical analysis is also needed to clarify how the
multiplier-dependent continuous free-energy term affects phase structure,
constraint satisfaction, and the number of useful low-energy states.
Replica methods and related typical-case analyses provide a natural route
to this question.

Beyond QpMP, relevant application domains include congested facility location~\cite{FISCHETTI2016557}, thermal unit commitment~\cite{YANG2015195_UC}, and cardinality-constrained regression~\cite{bertsimas2016best}, which combine binary decisions with continuous quantities and quadratic costs. 

As discussed in Appendix~\ref{app:regularity}, the continuous sector need not be restricted to the linear nonnegative form considered in the QpMP experiments; the essential requirement is that its log-partition function and multiplier derivatives remain tractable. This includes Gaussian and componentwise separable sectors and can, in principle, extend to more general nonlinear or nonconvex continuous sectors when the corresponding integrals can still be evaluated efficiently. On the binary side, preserving a QUBO inner problem requires the
multiplier-dependent binary Hamiltonian to remain at most quadratic; higher-order binary terms would require an additional quadratization step.

These observations suggest two complementary directions for future work. One is to identify reformulations that bring a broader class of continuous sectors into this analytically tractable setting. The other is to extend the framework to sectors whose log-partition functions are not available in closed form, including bounded or polyhedrally constrained domains.

More broadly, the present viewpoint may also be useful for decomposition
methods in which a QUBO sampler is used inside a multiplier-driven
subproblem. For example, it may help characterize annealing-based pricing
in Dantzig--Wolfe decomposition and column generation
~\cite{Hirama_jpsj_2023,takabayashi_2025_cg}, particularly the quality,
robustness, and diversity of generated columns.

\appendix
\titleformat{\section}
  {\color{color1}\large\sffamily\bfseries}
  {Appendix~\thesection.}
  {0.5em}
  {#1}
  []
\section{Regularity conditions and tractable continuous-variable integrals}
\label{app:regularity}
This appendix gives sufficient conditions for interchanging differentiation
and integration in Eq.~\eqref{eq:iy_def}, with inverse temperature
$\beta>0$ fixed.
 Define
\begin{equation}
\mathcal{D}
=
\left\{
\bm{\lambda}\in\R^L
\mid
I_y(\bm{\lambda})<\infty
\right\}.
\end{equation}
For any $\bm{\lambda}_0\in\mathcal{D}$, assume that there is a neighborhood $U\subset\mathcal{D}$ and integrable functions $M_0(\bm{y})$ and $M_k(\bm{y})$, $k=1,\ldots,L$, such that, for all $\bm{\lambda}\in U$,
\begin{equation}
\exp\left\{
-\beta\left[
g_0(\bm{y})-\bm{\lambda}^{\top}\bm{G}(\bm{y})
\right]
\right\}
\le
M_0(\bm{y}),
\label{eq:dominating_0}
\end{equation}
and
\begin{equation}
\left|G_k(\bm{y})\right|
\exp\left\{
-\beta\left[
g_0(\bm{y})-\bm{\lambda}^{\top}\bm{G}(\bm{y})
\right]
\right\}
\le
M_k(\bm{y}),
\quad k=1,\ldots,L.
\label{eq:dominating_k}
\end{equation}
The dominated convergence theorem then implies that $I_y(\bm{\lambda})$ is differentiable on $U$ and that the derivative may be passed through the integral. Hence
\begin{align}
\frac{\partial I_y(\bm{\lambda})}{\partial\lambda_k}
 = 
\int_Y d\bm{y}\,
\beta G_k(\bm{y})
\exp\left\{
-\beta\left[
g_0(\bm{y})-\bm{\lambda}^{\top}\bm{G}(\bm{y})
\right]
\right\}
\nonumber = 
\beta I_y(\bm{\lambda})
\left\langle G_k(\bm{y})\right\rangle_{P_{\bm{\lambda}}}.
\end{align}
Therefore,
\begin{equation}
\frac{\partial}{\partial\lambda_k}
\left[-\frac{1}{\beta}\log I_y(\bm{\lambda})\right]
=
-\left\langle G_k(\bm{y})\right\rangle_{P_{\bm{\lambda}}}.
\label{eq:derivative_logIy_appendix}
\end{equation}

By contrast, the binary contribution is a finite sum, so
\begin{equation}
\frac{\partial}{\partial\lambda_k}
\left[-\frac{1}{\beta}\log Z_x(\bm{\lambda})\right]
=
-\left\langle F_k(\bm{x})\right\rangle_{Q_{\bm{\lambda}}}.
\label{eq:derivative_logZx_appendix}
\end{equation}
Combining Eqs.~\eqref{eq:derivative_logIy_appendix} and
\eqref{eq:derivative_logZx_appendix} gives Eq.~\eqref{eq:multiplier_residual}.

These conditions apply to the general continuous integral. In the MBQP case,
the derivative can instead be verified directly from the closed-form
expression in Eq.~\eqref{eq:mbqp_iy_closed} considered in the main text.

The key requirement of the broader formulation is the tractability of
\eqref{eq:iy_def} and its multiplier derivatives. Thus, the approach is not
limited to the linear nonnegative continuous sector used in the MBQP~\eqref{eq:mbqp}.
 For example, if $G_k(\bm{y})=\bm{c}_k^{\top}\bm{y}$ and $g_0(\bm{y}) = \frac{1}{2}\bm{y}^{\top}P\bm{y} + \bm{q}^{\top}\bm{y}$ with $P\succ0$ and
$Y=\mathbb{R}^{M}$, then $I_y$ is a Gaussian integral and
\begin{equation}
I_y(\bm{\lambda})
=
\left(\frac{2\pi}{\beta}\right)^{M/2}
(\det P)^{-1/2}
\exp\left\{
\frac{\beta}{2}
(C\bm{\lambda}-\bm{q})^{\top}
P^{-1}
(C\bm{\lambda}-\bm{q})
\right\},
\label{eq:gaussian_integral}
\end{equation}
where $C=(\bm{c}_1,\ldots,\bm{c}_L)$. Hence,
$\langle\bm{y}\rangle=P^{-1}(C\bm{\lambda}-\bm{q})$, and the continuous
gradient contribution is available in closed form. 

More generally, the continuous sector enters the proposed framework only
through the log-partition function $I_y(\bm{\lambda})$ and its derivatives
with respect to the multipliers. Therefore, the practical scope of the
formulation is determined primarily by whether these quantities can be
evaluated analytically or at sufficiently low computational cost, rather
than by the particular linear form used in the MBQP example.
The Gaussian and componentwise separable cases above illustrate two classes for which this requirement can be satisfied.

This observation also allows the formulation, in principle, to accommodate more general nonlinear continuous sectors and, when the corresponding integrals remain well defined and tractable, nonconvex ones. For such problems, however, closed-form evaluation is not guaranteed, and additional approximation or numerical integration may be required.

\section{Covariance curvature and the update rule}
\label{app:bb_stepsize}
This appendix explains the curvature underlying the projected BB multiplier
update. We use the residual defined in Eq.~\eqref{eq:multiplier_residual};
the multiplier condition is therefore
$\bm r(\bm{\lambda})=\bm 0$.

Under the regularity conditions of Appendix~\ref{app:regularity}, the
positive-semidefinite curvature of the dual potential is
\begin{equation}
K(\bm{\lambda})
=
-\nabla_{\bm{\lambda}}^2\Phi_\beta(\bm{\lambda}),
\end{equation}
whose components are
\begin{equation}
K_{k\ell}(\bm{\lambda})
=
\beta
\Cov_{Q_{\bm{\lambda}}}
\left[
F_k(\bm{x}),F_\ell(\bm{x})
\right]
\nonumber +
\beta
\Cov_{P_{\bm{\lambda}}}
\left[
G_k(\bm{y}),G_\ell(\bm{y})
\right].
\label{eq:curvature_cov_full}
\end{equation}
Thus, $\Phi_\beta$ is concave in the multiplier variables. If the full curvature matrix were used,
the Newton step for the stationarity condition $\bm r(\bm\lambda)=0$ would be
\begin{equation}
\bm{\lambda}^{(t+1)}
=
\bm{\lambda}^{(t)}
+
K(\bm{\lambda}^{(t)})^{-1}
\bm r(\bm{\lambda}^{(t)}).
\label{eq:newton_lambda_appendix}
\end{equation}
Indeed, since $\bm r(\bm \lambda)=\nabla_\lambda \Phi_\beta(\bm \lambda)$ and
$\nabla_\lambda \bm r(\bm \lambda)=\nabla_\lambda^2\Phi_\beta(\bm \lambda)=-K(\bm \lambda)$,
Newton's method for $\bm r(\bm \lambda)=0$ gives the above update. In practice, however,
applying this step would require estimating and inverting the full covariance curvature matrix.

For the MBQP case of the main text, which has a linear continuous sector,
the continuous covariance is also available in closed form. Since $y_j$ is exponentially distributed with rate $\beta d_j(\bm{\lambda})$,
\begin{equation}
\operatorname{Var}_{P_{\bm{\lambda}}}[y_j]
=
\frac{1}{\beta^2 d_j(\bm{\lambda})^2}.
\end{equation}
Consequently,
\begin{equation}
\beta
\Cov_{P_{\bm{\lambda}}}
\left[G_k(\bm{y}),G_{\ell}(\bm{y})\right]
=
\frac{1}{\beta}
\sum_{j=1}^{M}
\frac{V_{kj}V_{\ell j}}
{d_j(\bm{\lambda})^2}.
\label{eq:continuous_cov_closed}
\end{equation}
This expression diverges as $d_j(\bm{\lambda})\to0$, consistent with the increasing curvature of the logarithmic barrier near the effective-domain boundary.

In practice, the empirical covariance estimated from QUBO samples can be
ill-conditioned or rank-deficient, and the analytic continuous curvature
can become large near the boundary $d_j(\bm{\lambda})=0$. We therefore
use a Barzilai--Borwein (BB) step size~\cite{BARZILAI_BORWEIN1988}, which replaces the full curvature matrix by a scalar
secant approximation. Locally, $\bm r^{(t)}-\bm r^{(t-1)}\simeq -K_t(\bm\lambda^{(t)}-\bm\lambda^{(t-1)})$. Approximating $K_t^{-1}$ by $\alpha_t I$ gives the secant relation $\bm\lambda^{(t)}-\bm\lambda^{(t-1)}\simeq-\alpha_t(\bm r^{(t)}-\bm r^{(t-1)})$. The least-squares scalar is Eq.~\eqref{eq:BB_stepsize}.

Given $\alpha_t$, the projected multiplier update is the quadratic program
\begin{equation}
\bm{\lambda}^{(t+1)}
=
\argmin_{\bm{\lambda}\in\mathcal{D}_{\varepsilon}}
\left\{
-
(\bm r^{(t)})^{\top}
(\bm{\lambda}-\bm{\lambda}^{(t)})
+
\frac{1}{2\alpha_t}
\|\bm{\lambda}-\bm{\lambda}^{(t)}\|_2^2
\right\}.
\label{eq:projected_bb_qp}
\end{equation}
Without the domain constraint, Eq.~\eqref{eq:projected_bb_qp} reduces to $\bm{\lambda}^{(t+1)} = \bm{\lambda}^{(t)}+\alpha_t\bm{r}^{(t)}$. Thus, the projected BB update approximates the inverse covariance curvature by a scalar secant step while projection preserves the effective domain, avoiding explicit estimation and inversion of the full curvature matrix.

\section{Runtime breakdown of the proposed method}
\label{app:runtime_breakdown}

To identify the main computational costs of the proposed method, we decompose the elapsed time for the representative target $\epsilon=10^{-2}$ into QUBO sampling, multiplier-residual computation, and all remaining computations.

\begin{figure}[H]
\centering
\includegraphics[width=0.7\textwidth]{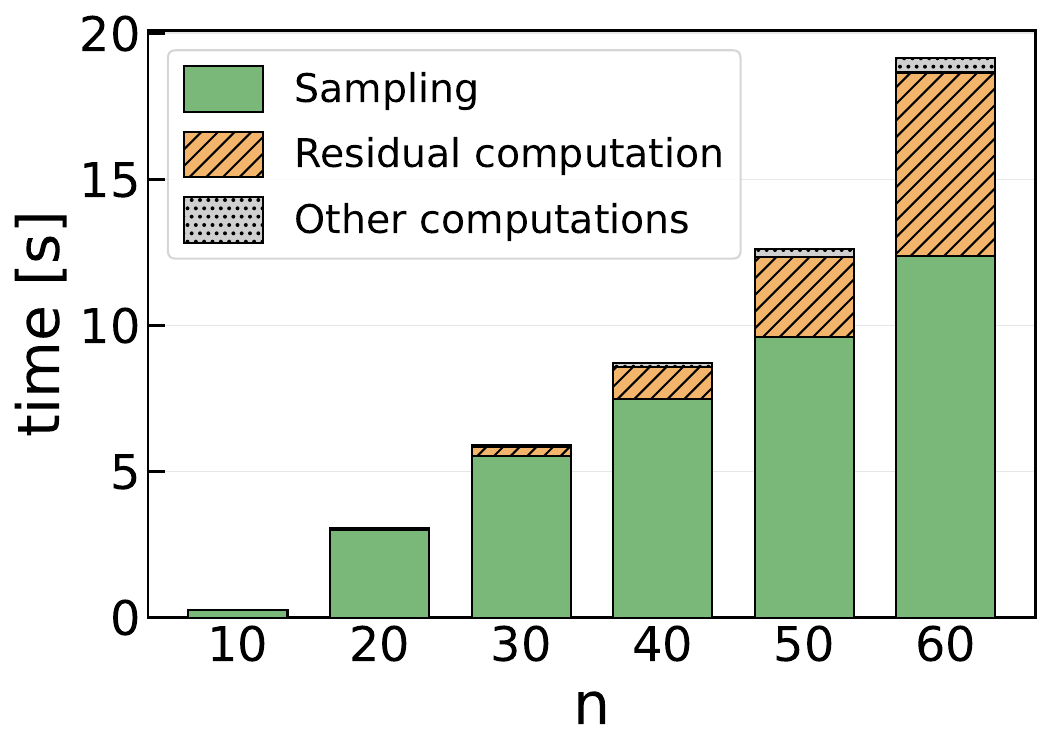}
\caption{Runtime breakdown of the proposed method for the fixed
relative-error target $\epsilon=10^{-2}$. The QpMP instances use
$\rho=0.1$, $p=0.2n$, and $\beta=0.1$. The bars show the mean elapsed
time over target-reaching runs, decomposed into QUBO sampling, multiplier-residual computation, and all remaining computations. The same stopping criterion and computational budget as in
Fig.~\ref{fig:ttt_fixed_error} are used.}
\label{fig:runtime_breakdown}
\end{figure}

Figure~\ref{fig:runtime_breakdown} shows the breakdown for
$\rho=0.1$, $p=0.2n$, and $\beta=0.1$, under the same settings as the
time-to-target experiment with prescribed relative-error targets.
QUBO sampling is the largest component over the tested range.
The cost of the multiplier-residual computation also increases with problem size, whereas the remaining computations account for only a small fraction of the total runtime.

\bibliography{main}

\section*{Acknowledgments}
We received financial support from the Cross-ministerial
Strategic Innovation Promotion Program (SIP) of the Cabinet
Office (No. 23836436).

\section*{Author contributions}
T.T. conceived of the presented idea and performed the experiments. M.O. verified the analytical methods and supervised the project. All authors discussed the results and contributed to the final manuscript.
\section*{Additional Information}
\subsection*{Data Availability}
No external datasets were used in this study.
All QpMP instances were generated synthetically using the procedure described in the Experimental settings and evaluation metrics section.
For each parameter setting, the instance index was used as the random seed, allowing the instances used in the numerical experiments to be reproduced from the source code provided in the accompanying code repository.

\subsection*{Code Availability}
The source code used to implement the proposed method, generate the QpMP
instances, and reproduce the numerical experiments reported in this study
is available at Zenodo: \url{https://doi.org/10.5281/zenodo.22790625}.

\end{document}